\documentclass[prl,twocolumn,superscriptaddress]{revtex4}

\usepackage[dvips]{color}
\usepackage{amsmath}
\usepackage{graphicx}
\usepackage{amssymb}
\usepackage{amsthm, amscd}
\usepackage{amsfonts}
\usepackage{cancel}
\usepackage{times}
\usepackage{pstricks}
\usepackage{wasysym}
\usepackage{ulem}
\usepackage[dvips]{epsfig}

\newcommand{\dslash}{\not{\hbox{\kern-2pt $\partial$}}}

\begin{document}


\title{Particle-Hole Symmetry and the $\nu=\frac{5}{2}$
Quantum Hall State}

\author{Sung-Sik Lee}
\affiliation{Kavli Institute for Theoretical Physics,
University of California, Santa Barbara, CA 93106-4030}

\author{Shinsei Ryu}
\affiliation{Kavli Institute for Theoretical Physics,
University of California, Santa Barbara, CA 93106-4030}

\author{Chetan Nayak}
\affiliation{Microsoft Research, Station Q, CNSI Building,
University of California, Santa Barbara, CA 93106-4030}
\affiliation{Department of Physics and Astronomy,
University of California, Los Angeles, CA 90095-1547}

\author{Matthew P.A. Fisher}
\affiliation{Microsoft Research, Station Q, CNSI Building,
University of California, Santa Barbara, CA 93106-4030}

\date{\today}

\begin{abstract}
We discuss the implications of approximate particle-hole
symmetry in a half-filled Landau level in which a paired
quantum Hall state forms. We note that the Pfaffian state
is not particle-hole symmetric. Therefore, in the limit
of vanishing Landau level mixing, in which particle-hole
transformation is an exact symmetry, the Pfaffian spontaneously
breaks this symmetry. There is a particle-hole conjugate
state, which we call the anti-Pfaffian, which is degenerate
with the Pfaffian in this limit. We observe that strong Landau level
mixing should favor the Pfaffian, but it is an open problem
which state is favored for the moderate Landau level mixing
which is present in experiments. We discuss the bulk and edge
physics of the anti-Pfaffian. We analyze a simplified model
in which transitions between analogs of the two states can be
studied in detail. Finally, we discuss experimental implications.
\end{abstract}

\maketitle


\paragraph{Introduction.}
The nature of the observed quantum Hall plateau at
$\sigma_{xy}=\frac{5}{2}\frac{e^2}{h}$ \cite{Willett87,Xia04} is an important
unresolved problem. If the Moore-Read Pfaffian state
\cite{Moore91,Greiter92,Nayak96c,Read96}
were realized at this plateau, then it would be the first non-Abelian topological
phase observed in nature. Such a discovery
could pave the way towards the realization of a
topological quantum computer \cite{DasSarma05}.
However, although several experiments have been proposed
which could directly confirm it
\cite{Fradkin98,DasSarma05,Bonderson06,Stern06},
the only evidence which currently suggests that the
$\nu=5/2$ plateau is in the universality class of the
Pfaffian state is the numerical diagonalization of
the Hamiltonian for systems with a small number of electrons \cite{Morf98,Rezayi00}.
In this paper, we add a new wrinkle to this discussion.
We note that the particle-hole conjugate of the Pfaffian is
a new state which will be exactly degenerate in energy with
the Pfaffian in the limit of vanishing Landau-level mixing.
Landau level mixing is a symmetry-breaking perturbation
which lifts the degeneracy between the two states.
It is an open problem which state is favored at the moderate
Landau level mixing which is present in experiments. Since the
effects of Landau level mixing have not been fully accounted
for in numerics, we suggest that the anti-Pfaffian is, at the very least,
a new candidate for the observed $\nu=5/2$ state.

When Landau level mixing is neglected, the Hamiltonian
for electrons at filling fraction $\nu=N+\frac{p}{q}$ can be related
to one for electrons at $\nu=N+1-\frac{p}{q}$
by an anti-unitary particle-hole transformation,
${c_{m}^\dagger}\rightarrow{c_{m}}$,
${c_{m}}\rightarrow{c_{m}^\dagger}$, with $m$ labeling orbitals
within the Landau level. The Hamiltonian transforms according to
${H_2}\rightarrow {\tilde H}_{2} + \text{const.}$, where
\begin{multline}
\label{eqn:2-body-Ham}
{H_2}=\sum_{klmn} V_{klmn} {c_{k}^\dagger}{c^{}_m} {c_{l}^\dagger} {c^{}_n}
- \mu {\sum_m} {c_{m}^\dagger}{c^{}_m}\, ,\\
{\tilde H}_{2} = \sum_{klmn} V_{klmn} {c_{k}^\dagger}{c^{}_m}
{c_{l}^\dagger} {c^{}_n} +
\left(\mu- 2\mu_{1/2}\right){\sum_m} {c_{m}^\dagger}{c^{}_m}
.
\end{multline}
Here, $V_{klmn}$ are the matrix elements
of the Coulomb interaction between the corresponding orbital states,
and $\mu_{1/2}={\sum_n}V_{mnmn}$ (the sum is independent of $m$ in a
translationally-invariant system).
For the special case of $\frac{p}{q}=\frac{1}{2}$, for which
$\mu=\mu_{1/2}$, this is a symmetry of the system.

It is widely believed, on the basis of numerical evidence
\cite{Morf98,Rezayi00} that the experimentally-observed plateau
at $\nu=\frac{5}{2}$ is in the universality class of the
Moore-Read Pfaffian state, by which it is meant that
the lowest Landau level (of both spins) is filled, and the
electrons in the first excited Landau level are fully spin-polarized
and have a wavefunction which is in the same universality
class as the one given by acting with Landau level raising
operators on:
\begin{equation}
\label{eqn:Pfaffian-wvfn}
    \Psi_{\rm Pf} ({z_i})
   = \text{Pf}\!\left(\frac{1}{z_i - z_j}\right)\,
   \prod_{i<j}(z_i - z_j)^{2}\,
   e^{-\sum {\left|{z_i}\right|^2}/4{\ell_0^2}}
.
\end{equation}
However, this state is not invariant under a particle-hole
transformation. Let $\sigma_{xy}$
and $\kappa_{xy}$ denote the contributions to the electrical and thermal
Hall conductivities (in units of ${e^2}/h$ and ${\pi^2}{k_B^2}T/3h$,
respectively) of the fractional state in the lowest
unfilled Landau level, in which the Pfaffian is assumed to
form (which, in the case of $\nu=5/2$, is the first excited
Landau level). Under a particle-hole transformation
of this Landau-level,
$\sigma_{xy}\rightarrow 1-\sigma_{xy}$
and 
$\kappa_{xy}\rightarrow 1-\kappa_{xy}$.
The Pfaffian state has 
$\kappa_{xy}= \frac{3}{2}$,
as may be seen most easily from its edge theory \cite{Milovanovic96},
which has two modes, a chiral boson $\phi$ and a chiral Majorana
fermion $\psi$,  propagating in the same direction:
\begin{equation}
{\cal L}_\text{Pf}(\psi,\phi) = {\psi}(-{\partial_t}+i{v_n}\partial_x){\psi}
+ \frac{2}{4 \pi} {\partial_x}{\phi}(i{\partial_t}+{v_c}{\partial_x}){\phi} 
.
\label{eqn:Pf-edge}
\end{equation}
Here, $e^{i\phi}$ creates a charge $e/2$ semion with scaling dimension
$1/4$, and $v_c, v_n$ denote the two edge velocities.
A chiral boson  contributes $\kappa_{xy}= 1$, while
a chiral Majorana fermion carries $\kappa_{xy}= \frac{1}{2}$.
The particle-hole conjugate of the Pfaffian, which we
will call the anti-Pfaffian ($\overline{\text{Pf}}$), with
wavefunction of the form:
\begin{multline}
\label{eqn:anti-Pfaff-wvfn}
\Psi_{\overline{\text{Pf}}}= \int\mbox{$\prod_\alpha$}{d^2}{\eta_\alpha}\,\,
 \mbox{$\prod_{i<j}$}\left({z_i} - {z_j}\right)\,
 e^{-\sum {\left|{z_i}\right|^2}/4{\ell_0^2}}\,\times\\
\mbox{$\prod_{i,\alpha}$}\left({z_i}-{\eta_\alpha}\right)\,
\mbox{$\prod_{\beta<\gamma}$}\left({\eta_\beta} - {\eta_\gamma}\right)\,
e^{-\sum {\left|{\eta_\alpha}\right|^2}/4{\ell_0^2}}
\Psi_{\rm Pf}(\overline{\eta_\alpha})
\end{multline}
must have $\kappa_{xy}=- \frac{1}{2}, \sigma_{xy} = \frac{1}{2}$.
Therefore, the anti-Pfaffian has counter-propagating edge modes,
which will have direct experimental significance, as we discuss later.

Landau level mixing breaks particle-hole symmetry. If we treat
it perturbatively (although it is not particularly small in experiments)
then, in addition to renormalizing the Coulomb repulsion,
which does not break particle-hole symmetry, it also generates
three-body interactions, which do break the symmetry.
In fact, the Pfaffian (\ref{eqn:Pfaffian-wvfn}) is the exact ground state
of the simplest repulsive non-vanishing three-body interaction.
The anti-Pfaffian (\ref{eqn:anti-Pfaff-wvfn}) is, therefore, the exact
ground state of the particle-hole conjugate of this interaction,
which is an attractive three-body interaction together with a repulsive two-body
interaction with the same coefficient. When two-body Coulomb
interactions and weaker three-body interactions are present, it
is unclear which phase occurs. (Although we expect that the
specific wavefunction (\ref{eqn:Pfaffian-wvfn}) is lower in energy
than (\ref{eqn:anti-Pfaff-wvfn}) if the three-body interaction is
repulsive, this does not tell us which phase the actual ground state
of the Hamiltonian is in).
Fermion Chern-Simons theory,
at the mean-field level, which includes non-perturbative
mixing between all Landau levels, implies that the
Pfaffian is the ground state \cite{Greiter92}. However,
it is an open question which state is favored by the moderate
Landau level mixing which actually occurs in experiments,
where the strength of Landau level mixing,
$\left({e^2}/{\epsilon{\ell_0}}\right)/\hbar{\omega_c}\sim 1$.
On the sphere, which does not have translational symmetry,
a particle-hole transformation is not a symmetry of (\ref{eqn:2-body-Ham});
the Pfaffian and anti-Pfaffian states of $N$ particles occur at magnetic fluxes
of ${N_\Phi}=2N-3$ and ${N_\Phi}=2N+1$, respectively. This may
explain why the anti-Pfaffian has not been identified in numerical
studies of finite systems on the sphere \cite{Morf98}. A particle-hole transformation
is a symmetry of (\ref{eqn:2-body-Ham}) on the torus. However, in a finite
system, we expect mixing between the Pfaffian and the anti-Pfaffian
so that the symmetric combination is the ground state, consistent
with numerics \cite{Rezayi00}.

\paragraph{Edge Excitations of the anti-Pfaffian}
Under a particle-hole transformation, the edge between the anti-Pfaffian
and the vacuum ($\nu=0$) is mapped onto the edge between
the Pfaffian and a $\nu=1$ Hall liquid. (At $\nu=5/2$, there are
also the edges of the lowest Landau level of both spins,
but they play no role in our discussion, so we drop them for simplicity.)
Therefore, we can deduce
the former from the properties of the latter, in which
the edge of the Pfaffian (\ref{eqn:Pf-edge}) is coupled to a
counter-propagating free chiral Dirac fermion (or its bosonized
equivalent):
\begin{multline}
{\cal L} =  \frac{1}{4 \pi} {\partial_x}{\phi_1}(-i{\partial_t}+v_{1}{\partial_x}){\phi_1}
+ {\cal L}_\text{Pf}({\psi_1},{\phi_2})\\
 + \frac{1}{4\pi} 2v_{12}{\partial_x}{\phi_1}{\partial_x}{\phi_2}
 + \xi(x)\, {\psi_1}\, e^{i({\phi_1}-2{\phi_2})} + \text{h.c.} .
\end{multline}
We have included a density-density interaction between
the $\nu=1$ and $ \nu=1/2$  edge modes, and also an inter-mode electron tunneling
term.  
With an assumed inter-mode momentum mismatch 
and in the presence of impurities, the electron tunneling amplitude $\xi(x)$ can be taken as a random (complex) function
with zero mean and short-ranged correlations,
$\overline{\xi^*(x)\xi(x')}=W\delta(x-x')$. 
For large $v_{12}$ the tunneling term is relevant,
and can then be conveniently analyzed\cite{Kane94}
by introducing a charge/neutral decomposition, $\phi_\rho = \phi_1 - \phi_2$ and 
$\phi_{\sigma}={\phi_1}-2{\phi_2}$,  and then fermionizing the neutral chiral boson:
 $e^{i\phi_{\sigma}}\equiv {\psi_2}+i{\psi_3}$.
The Lagrangian then takes the form ${\cal L}={\cal L}_\text{sym}
+ {\cal L}_\text{pert}$, where:
\begin{equation}
\label{eqn:edge-action-sym}
{\cal L}_\text{sym} = \frac{2}{4 \pi} {\partial_x}{\phi_\rho}
(-i{\partial_t}+v_{\rho}{\partial_x}){\phi_\rho}
+ {\psi_a}(-{\partial_t}+i
v_{\sigma}{\partial_x}){\psi_a}
\end{equation}
\begin{equation}
\label{eqn:edge-action-pert}
{\cal L}_\text{pert} = 
2 i {\psi_1}( \xi_1 {\psi_3} 
+ {\xi_2}{\psi_2})
+ \delta v_{1}{\psi_1}i{\partial_x}{\psi_1}
+ i v {\psi_2}{\psi_3}\, {\partial_x}{\phi_\rho}
\end{equation}
with ${\xi_1}$, ${\xi_2}$ the real and imaginary
parts of $\xi(x)$.
The three Majorana fermions $\psi_a$, $a=1,2,3$
form an SU(2)$_2$ triplet in the absence of the symmetry-breaking
terms (\ref{eqn:edge-action-pert}).
The first terms in (\ref{eqn:edge-action-pert}) can be eliminated
from the action by an SU(2) rotation,
$\psi = O\tilde{\psi}$, with
$O(x)={\text{P}}\exp\left(-i{\int_{-\infty}^x} dx'\,
({\xi_1}(x'){T_2} - {\xi_2}(x'){T_3})/{v_\sigma}\right)$,
where $P$ denotes a path ordering of the integral, and
$T_a$, $a=1,2,3$ are the SU(2) generators in the spin-1
representation. In the transformed variables, 
the second and third terms in (\ref{eqn:edge-action-pert})
will have spatially dependent random coefficients.
Integrating out $\xi_1$, $\xi_2$ (using the replica method, for
instance), results in terms which have scaling dimension $(-1)$
and are, therefore, perturbatively irrelevant.
Hence, we obtain  (\ref{eqn:edge-action-sym})
as the action for the edge between the Pfaffian state and
a $\nu=1$ Hall droplet.
The edge theory between the anti-Pfaffian state and the vacuum
is obtained by flipping the directions of all the modes in (\ref{eqn:edge-action-sym}).
There are {\it three} counter-propagating neutral Majorana
fermion edge modes, which yields the expected value
for the thermal Hall conductivity, 
$\kappa_{xy}=-1/2$.

In the anti-Pfaffian edge theory, the minimal dimension electron operator is $e^{i\phi_1}=({\psi_2}-i{\psi_3})\,e^{2i\phi_\rho}$.
There are $6$ primary fields which are local with respect
to this electron operator: $1$, $e^{i\phi_\rho}$, ${\psi_a}$,
${\psi_a}e^{i\phi_\rho}$, $\phi_{1/2}^{\alpha}\,e^{i{\phi_\rho}/2}$,
$\phi_{1/2}^{\alpha}e^{3i{\phi_\rho}/2}$.  The spin-$1/2$ 
primary fields of SU(2)$_2$, denoted $\phi^\alpha_{1/2}$,  can be written in
terms of the Ising spin and disorder fields $\sigma_a$ and $\mu_a$
of the three Majorana fermions: $\phi_{1/2}^{+}={\sigma_1}{\sigma_2}{\sigma_3}
+ i{\mu_1}{\mu_2}{\sigma_3} + {\mu_1}{\mu_2}{\mu_3} +
i{\sigma_1}{\sigma_2}{\mu_3}$ and
$\phi_{1/2}^{-}={\sigma_1}{\mu_2}{\mu_3}
- i{\mu_1}{\sigma_2}{\sigma_3} - i{\sigma_1}{\mu_2}{\sigma_3} +
{\mu_1}{\sigma_2}{\sigma_3}$.
The fields $\phi_{1/2}^{\alpha}$ thus act to switch between periodic and anti-periodic boundary conditions
on all three Majorana fermions  .
Note that $\phi_{1/2}$ is a dimension
$3/16$ operator, unlike $\sigma$, which has dimension $1/16$.
The difference in scaling dimension has consequences for quasiparticle
tunneling as we discuss below.

\paragraph{Topological Properties of the anti-Pfaffian}
The $6$ primary fields of the conformal field theory for the
anti-Pfaffian correspond to its $6$-fold degeneracy on the torus.
A $2d$ (bulk) effective field theory for the anti-Pfaffian state
which encodes this degeneracy as well as the other bulk topological properties,
can be deduced from consistency with the edge
theory or, alternatively, in the following way.
We begin by bosonizing the action for electrons at $\nu=1/2$ 
employing a Chern-Simons gauge field $c_\mu$:
\begin{multline}
{\cal L} = {\Phi^*}\left(i{\partial_0}+{c_0}+{A_0}\right)\Phi +
\frac{1}{2m}{\left|\left(i{\partial_i}+{c_i}+{A_i}\right)\Phi\right|^2}\\ + V(\Phi)
+ \frac{1}{4\pi}
 {c_\mu}\epsilon_{\mu\nu\lambda}{\partial_\nu}{c_\lambda}
,
\end{multline}
where $A_\mu$ is the electromagnetic field.
At the mean-field level, this is now a system of bosons at $\nu=1$.
We next use boson-vortex duality\cite{DHL_MPAF} to transform to a system
of vortices, also at $\nu_\text{eff}=1$, minimally coupled
to a dual gauge field $b_\mu$.   Integrating out $c_\mu$
induces a Chern-Simons term
for the $b_\mu$:
\begin{multline}
{\cal L} = {\Phi_v^*}\left(i{\partial_0}+{b_0}\right){\Phi_v} +
\frac{1}{2m_v}{\left|\left(i
{\partial_i}+{b_i}\right){\Phi_v}\right|^2}
+ U({\Phi_v})\\
- \frac{1}{4\pi} {b_\mu}\epsilon_{\mu\nu\lambda}{\partial_\nu}{b_\lambda}
-\frac{1}{2\pi}{b_\mu}\epsilon_{\mu\nu\lambda}{\partial_\nu}{A_\mu}
.
\end{multline}
We now shift ${b_\mu}\rightarrow{b_\mu}-{A_\mu}$ and refermionize
${\Phi_v}$ since $b_\mu$ attaches one flux quantum to it:
\begin{multline}
{\cal L} = \overline{\chi}\left(i{\partial_0}-{A_0}-\mu\right)\chi +
\frac{1}{2m_f}{\left|\left(i{\partial_i}-{A_i}\right)\chi\right|^2}\\ + 
\frac{1}{4\pi} {A_\mu}\epsilon_{\mu\nu\lambda}{\partial_\nu}{A_\lambda}
+ \ldots
\label{eqn:bulk-hole-action}
\end{multline}

 The fermion field $\chi^\dagger$ creates holes which carry electrical charge 
opposite to that of the electron. Therefore, when we attach flux to these
fermions, pair them, and condense the pairs \cite{Greiter92,Read00},
we will obtain a {\it negative}
contribution to the Hall conductivity. Furthermore, if the pairing
is due to gauge field fluctuations, as in \cite{Greiter92}, 
the pairs will have a $p-ip$ pairing symmetry.  The resulting
effective action can be written:
\begin{multline}
\label{eqn:holes-action}
{\cal L} = \overline{\eta} i\!\dslash \eta - m \overline{\eta} \eta
+ \frac{2}{4\pi} {a_\mu}\epsilon_{\mu\nu\lambda}{\partial_\nu}{a_\lambda}
- \frac{1}{4\pi} {{\tilde a}_\mu}\epsilon_{\mu\nu\lambda}{\partial_\nu}
{{\tilde a}_\lambda}\\
-  \frac{1}{2\pi}{A_\mu}\epsilon_{\mu\nu\lambda}{\partial_\nu}{a_\lambda}
+  \frac{1}{2\pi}{A_\mu}\epsilon_{\mu\nu\lambda}{\partial_\nu}{{\tilde a}_\lambda}
.
\end{multline}
The negative sign for the mass of the $2d$ Majorana fermion $\eta$ is a consequence
of the reversed pairing symmetry. The last term in (\ref{eqn:bulk-hole-action})
has been rewritten by introducing an auxiliary gauge field ${\tilde a}_\mu$.

Finally, we introduce
$a^{\scriptscriptstyle \rho}_{\mu} = {a_\mu}-{{\tilde a}_\mu}$
and $a^{\scriptscriptstyle \sigma}_{\mu} = 2{a_\mu}-{{\tilde a}_\mu}$
to obtain the desired $2+1d$ effective action for the anti-Pfaffian:
\begin{multline}
\label{eqn:anti-Pfaffian-bulk}
{\cal L} = \overline{\eta} i\!\dslash \eta - m \overline{\eta}\eta
- \frac{2}{4\pi} {a^{\scriptscriptstyle \rho}_\mu}
\epsilon_{\mu\nu\lambda}{\partial_\nu}{a^{\scriptscriptstyle \rho}_\lambda}
+ \frac{1}{4\pi} {a^{\scriptscriptstyle \sigma}_\mu}
\epsilon_{\mu\nu\lambda}{\partial_\nu}{a^{\scriptscriptstyle \sigma}_\lambda}\\
-  \frac{1}{2\pi}{A_\mu}\epsilon_{\mu\nu\lambda}{\partial_\nu}
a^{\scriptscriptstyle \rho}_{\mu}
.
\end{multline}
The action (\ref{eqn:anti-Pfaffian-bulk}) corresponds closely to the
edge effective action (\ref{eqn:edge-action-sym}). The most salient
topological feature is the reversed chirality of the neutral fermion
sector. As a result, the braid matrices are complex conjugated
as compared to the Pfaffian state \cite{Nayak96c,Ivanov01}:
$T_{ij} = e^{\frac{\pi i}{4}}\,e^{\frac{\pi}{4}{\gamma_i} {\gamma_j}}$.

\paragraph{Toy Model}
We next describe a simple {\it lattice} model of
spinless fermions  which has similar physics
to the neutral sectors of the Pfaffian and anti-Pfaffian phases.
The fermions hop only between near neighbor sites of a square lattice.
At half-filling the
model is invariant under the
anti-unitary symmetry ${c^{}_i}\rightarrow {(-1)^i}{c^\dagger_i}$,
${c^\dagger_i}\rightarrow {(-1)^i}{c^{}_i}$.
With interactions present we suppose that the fermions
can develop the following order parameters:
$\Delta_{ij}$, which is a $\Delta(p)=\sin{p_x}+i\sin{p_y}$ superconducting
order parameter which spontaneously breaks U(1);
$\varphi$, which spontaneously breaks particle-hole
symmetry by enabling next-nearest-neighbor hopping;
and $\theta$, which spontaneously breaks $\pi/2$
rotational symmetry. The mean-field Hamiltonian 
with these order parameters is:
\begin{multline*}
H = \sum_{\left\langle i,j\right\rangle} \left(-t{c^\dagger_i} {c^{}_j}
+ \Delta_{ij} {c^\dagger_i} {c^\dagger_j}  + \text{h.c.}\right)
-\mu {\sum_i} {c^\dagger_i} {c^{}_i}
\\
-\sum_{\left\langle\left\langle i,j\right\rangle\right\rangle}
 \varphi \,{c^\dagger_i} {c^{}_j} + \text{h.c.}
+  \sum_{i} \theta \left({c^\dagger_i} c^{}_{i+a\hat{x}}
- {c^\dagger_i} c^{}_{i+a\hat{y}}\right) + \text{h.c.}
\end{multline*}

When $\mu=\varphi=0$, this Hamiltonian is particle-hole symmetric.
When $\varphi=\theta=\mu = 0$, there are gapless excitations
at two ``nodes", with momenta ${\bf k}=(\pi,0), (0,\pi)$.
Such gapless excitations would not be present for an off-lattice $p+ip$
superconductor.  The low-energy excitations in the vicinity
of each nodes is a (two-component) Majorana fermion, which can be combined
into a single Dirac fermion $\psi$.
Non-zero $\varphi$ and $\theta$ are, respectively,
Dirac and Majorana mass terms for $\psi$.
The former breaks particle-hole symmetry, but not the O(2) which
rotates one Majorana fermion into the other (a $\pi/2$ lattice rotation
is $\pi$ rotation within this O(2)); the latter does not break
particle-hole symmetry but breaks O(2).

Let us suppose that $\varphi$ orders. Then $\mu$ must be adjusted
to maintain half-filling.
If $\varphi>0$, we need $\mu>0$, and
the electron Fermi surface (when $\Delta  = 0$)
is closed and the hole Fermi surface is open. The system is adiabatically
connect to electrons in the continuum, which essentially
form a continuum $p+ip$ superconductor.
In particular, it supports a gapless chiral Majorana fermion edge
mode \cite{Read00} with $\kappa_{xy}=\frac{1}{2}$. The long-distance form
of the pair wavefunction is $g({\bf r})\sim 1/z$. Hence, we identify this
phase with the neutral sector of the Pfaffian.  For  $\varphi<0$ 
the masses of the two nodal Majorana fermions change sign, 
giving $\kappa_{xy}=-\frac{1}{2}$.
In this case
$\mu<0$, and the holes form a closed
Fermi surface rather than the electrons.
The $\sin{p_x}+i\sin{p_y}$ pairs of electrons are, in this phase,
better interpreted as $\sin{p_x}-i\sin{p_y}$ pairs
of holes. This is consistent with our effective
field theory of the anti-Pfaffian (\ref{eqn:anti-Pfaffian-bulk}),
which has $p-ip$ pairs of composite
fermions obtained by attaching flux to holes.
Moreover, $g({\bf r})\sim {1}/{z} - {(-1)^x}/{\overline z} -
{(-1)^y}/{\overline z}$. We identify this phase with the neutral
sector of the anti-Pfaffian.

At the transition between these two phases,
$\varphi = 0$, and there are gapless excitations
described by a single massless  Dirac fermion:
\begin{equation}
{\cal L} = \overline{\psi} i\!\dslash \psi  - g {\left(\overline{\psi}\psi\right)^2} ,
\end{equation}
a $2d$ analogue of the Gross-Neveu model.
For $g<g_c$, the $\mathbb{Z}_2$
symmetry $\psi(t,x,y)\rightarrow {\gamma^1}\psi(t,-x,y)$,
$\overline{\psi}(t,x,y)\rightarrow 
\overline{\psi}(t,-x,y){\gamma^1}$
is unbroken ($\gamma^1$ is a purely imaginary $\gamma$ matrix),
and the nodal fermions are gapless.
However, for $g>g_c$, particle-hole symmetry can spontaneously break
$\varphi\propto\left\langle\overline{\psi}\psi\right\rangle\neq 0$,
and by varying explicit symmetry-breaking terms, such
as $\mu$ or a second-neighbor hopping $t'$, the system can  be driven through a first-order transition between
the two phases.
At $g=g_c$, the system is critical. Exponents are known in
the large-$N$ limit, where $N$ is the number of flavors of fermions,
e.g. $\nu=1+ 8/(3{\pi^2} N)$ \cite{Hands93}. As the critical
point is approached from within the symmery-broken phase,
the velocity of the gapless chiral Majorana fermion edge mode vanishes.
The small value of the neutral Majorana fermion edge mode
in numerical studies of the Pfaffian state may indicate
proximity to such a critical point \cite{Wan06}.
Note that with $\theta\neq 0$, there is also an intermediate phase
with $\kappa_{xy}=0$ which breaks rotational symmetry
(perhaps slightly reminiscent of the nematic phase at $\nu=5/2$
in a tilted field \cite{Pan99a}).

\paragraph{Transitions and Intermediate Phases}
Informed by the preceding discussion, we can write
down an effective field theory for the Pfaffian and anti-Pfaffian
states and the transition between them:
\begin{multline}
{\cal L} = \frac{-2}{4\pi} {a^{\scriptscriptstyle \rho}_\mu}
\epsilon_{\mu\nu\lambda}{\partial_\nu}{a^{\scriptscriptstyle \rho}_\lambda}
-  \frac{1}{2\pi}{A_\mu}\epsilon_{\mu\nu\lambda}{\partial_\nu}
a^{\scriptscriptstyle \rho}_{\mu}
+ \frac{1}{4\pi} {a^{\scriptscriptstyle \sigma}_\mu}
\epsilon_{\mu\nu\lambda}{\partial_\nu}{a^{\scriptscriptstyle \sigma}_\lambda}\\ +
{\Phi^*}\left(i{\partial_0}+{a^{\scriptscriptstyle \sigma}_0}\right){\Phi} +
\frac{1}{2m^*}{\left|\left(i
{\partial_i}+{a^{\scriptscriptstyle \sigma}_i}\right)
{\Phi}\right|^2} + U({\Phi})\\
+ \overline{\Psi} i\!\dslash \Psi - {g(\overline{\Psi}\Psi)^2}
.
\end{multline}
The phase in which $\left\langle\overline{\Psi}\Psi\right\rangle < 0$
and $\left\langle \Phi \right\rangle \neq 0$ corresponds to the Pfaffian state:
the gauge field ${a^{\scriptscriptstyle \sigma}_\mu}$ is gapped by the
Anderson-Higgs mechanism and the sign of the fermion $\Psi$'s mass
gives a right-handed Majorana fermion at the edge. The phase in
which $\left\langle\overline{\Psi}\Psi\right\rangle > 0$
and $\left\langle \Phi \right\rangle = 0$ corresponds to the anti-Pfaffian state:
there is a left-handed Dirac fermion at the edge associated with 
${a^{\scriptscriptstyle \sigma}_\mu}$ and a left-handed Majorana fermion
due to the sign of the mass of $\Psi$. At the critical point, both
$\Phi$ and $\Psi$ are critical. If $\Phi$ is fermionized using
${a^{\scriptscriptstyle \sigma}_\mu}$, then the critical theory has two
massless interacting Dirac fermions, in contrast to the toy model
which possessed only one, having no bosonic edge modes.

However, one can imagine a scenario in which only one of these
fields becomes critical. Then the system will go into a phase with
$\kappa_{xy}=\frac{1}{2}$, such as the phase in which
$\left\langle\overline{\Psi}\Psi\right\rangle > 0$
and $\left\langle \Phi \right\rangle \neq 0$, in which there is only
a single neutral Majorana fermion, but it is left-moving. In the
phase with $\left\langle\overline{\Psi}\Psi\right\rangle < 0$
and $\left\langle \Phi \right\rangle = 0$, which also has
$\kappa_{xy}=\frac{1}{2}$, there are three neutral Majorana fermions;
two are right-moving and one is left-moving. Although these
intermediate phases are logical possibilities, they are not
related to the Pfaffian or anti-Pfaffian phases by particle-hole symmetry,
and may be much higher in energy.

\paragraph{Discussion}
Our investigations open the door to a number of interesting
questions. In the toy model, the difference between the two
phases can be understood in terms of Fermi surface topology.
Can the difference between the Pfaffian and anti-Pfaffian states
be understood in terms of similar momentum space structure
in the lowest Landau level? If the Pfaffian and anti-Pfaffian states
are analogous to the two ordered states of the Ising model
at low temperatures (spin-up and spin-down), then what is the
analogue of the high-temperature phase? In the toy model, it
is a critical superconducting phase with two gapless $(2+1)-d$ Majorana fermions.
In the quantum Hall context, is it an analogous phase with bulk gapless
neutral excitations yet with a quantized Hall conductance?
Finally, there is the question of which state (if any of these)
is seen in experiments.
Since the main arguments for the Pfaffian state derive
from numerical studies which do not account for the effects of
Landau level mixing, we suggest that they be reconsidered
in light of the anti-Pfaffian state. 
One signature of the Pfaffian and anti-Pfaffian phases are
their differing thermal conductivities, 
$\kappa_{xy}=3/2$
and 
$\kappa_{xy}=-1/2$, respectively.
Electrical transport measurements through a point contact
will also differ in the two phases.  As described in
Ref. \onlinecite{Fendley06}, weak tunneling of the charge $e/4$ non-Abelian quasiparticles  between
the edges of a Pfaffian Hall bar leads to $R_{xx}\sim T^{-3/2}$.
For the anti-Pfaffian one obtains $R_{xx}\sim T^{-1}$,
different due to the extra edge modes present.
(See Ref. \onlinecite{Miller07} for experiments
in this direction.)
Interestingly, weak inter-edge tunneling
of the charge $e/2$ Laughlin quasiparticle also gives
 $R_{xx}\sim T^{-1}$, in both the Pfaffian and anti-Pfaffian states.
The existence of counter-propagating neutral modes in the anti-Pfaffian state might also be
detectable, and would 
have implications for interferometry experiments
\cite{Fradkin98,DasSarma05,Stern06,Bonderson06} if it proves
to be realized at $\nu=5/2$.

Upon completion of this work, we became aware of
similar predictions made by Levin {\it et al.},
arXiv:0707.0483.
This research has been supported by the NSF
under grants  
PHY05-51164 (S.L. and S.R.),
DMR-0529399 (M.P.A.F.),
DMR-0411800 (C.N.), and by the ARO under grant W911NF-04-1-0236 (C.N.).

\vskip -0.5cm


\begin{thebibliography}{29}

\bibitem{Willett87}
R.L. Willett {\it et al.}, Phys. Rev. Lett. {\bf 59}, 1776 (1987).

\bibitem{Xia04}
J.S. Xia {\it et al.}, Phys. Rev. Lett. {\bf 93}, 176809 (2004);
J.P. Eisenstein {\it et al.}, \prl 88, 076801 (2002).

\bibitem{Moore91}
G. Moore and N. Read, Nucl. Phys. B {\bf 360},  362  (1991).

\bibitem{Greiter92}
M. Greiter {\it et al.}, Nucl. Phys. B {\bf 374},  567  (1992).

\bibitem{Nayak96c}
C. Nayak and F. Wilczek, Nucl. Phys. B {\bf 479},  529  (1996).

\bibitem{Read96}
N. Read and E. Rezayi, Phys. Rev. B {\bf 54},  16864  (1996).

\bibitem{Fradkin98}
E. Fradkin {\it et al.}, Nucl. Phys. B {\bf 516},  704 (1998).

\bibitem{Morf98}
R.~H. Morf, Phys. Rev. Lett. {\bf 80},  1505  (1998).

\bibitem{Rezayi00}
E.~H. Rezayi and F.~D.~M. Haldane, Phys. Rev. Lett. {\bf 84}, 4685 (2000).

\bibitem{DasSarma05} S. Das Sarma, M. Freedman, and C. Nayak,
\prl {\bf 94}, 166802 (2005).

\bibitem{Stern06} A. Stern  and B. I. Halperin, \prl {\bf 96}, 016802 (2006).

\bibitem{Bonderson06} P. Bonderson {\it et al.}, \prl {\bf 96}, 016803 (2006).

\bibitem{Milovanovic96} M. Milovanovi\'c and N. Read,
Phys. Rev. B {\bf 53}, 13559 (1996).

\bibitem{Kane94} C.~L. Kane, M.~P.~A. Fisher, and J. Polchinski,
Phys. Rev. Lett. {\bf 72}, 4129 (1994);
C~.L. Kane and M.~P.~A. Fisher,
Phys. Rev. B {\bf 51}, 13449 (1995).

\bibitem{Read00}
N. Read and D. Green, Phys. Rev. B {\bf 61}, 10267 (2000).

\bibitem{Ivanov01}
D.~A. Ivanov, Phys. Rev. Lett. {\bf 86},  268  (2001).

\bibitem{DHL_MPAF}
D.~H. Lee and M.~P.~A. Fisher, Phys. Rev. Lett. {\bf 63}, 903 (1989).

\bibitem{Hands93} S. Hands {\it et al.},
Ann. Phys. (N.Y.) {\bf 224}, 29 (1993).

\bibitem{Wan06} X. Wan {\it et al.}, \prl {\bf 97},
256804 (2006).

\bibitem{Pan99a} W. Pan {\it et al.}, \prl {\bf 83}, 820 (1999);
M.~P. Lilly {\it et al.}, \prl {\bf 83}, 824 (1999).

\bibitem{Fendley06}
P. Fendley, M.~P.~A. Fisher, and C. Nayak, Phys. Rev. Lett. {\bf 97}, 036801 (2006);
\prb {\bf 75}, 045317 (2007).

\bibitem{Miller07}
J.~B. Miller {\it et al.}, cond-mat/0703161.


\end{thebibliography}

\end{document}